\documentclass[aps,prd,twocolumn,groupedaddress,showpacs,showkeys,preprintnumbers]{revtex4}

\usepackage{amsmath,graphicx,natbib}

\newcommand{\B}[1]{\textbf{#1}}
\newcommand{\bsym}[1]{\boldsymbol{#1}}
\newcommand{\ew}{autosum}
\newcommand{\mix}{mixed}
\newcommand{\fullint}{\int_{-\infty}^{\infty}}
\newcommand{\halfint}{\int_{0}^{\infty}}
\newcommand{\matx}[1]{\overline{\textbf{#1}}}
\newcommand{\mc}[1]{\mathcal{#1}}
\newcommand{\var}{\text{var}}
\newcommand{\op}{\text{op}}
\newcommand{\wh}{\text{wh}}
\newcommand{\thsub}{\text{th}}
\newcommand{\boxsub}{\text{box}}
\newcommand{\trsub}{\text{tr}}
\newcommand{\ewsub}{\text{as}}
\newcommand{\bsub}{\text{B}}

\begin{document}

\preprint{FERMILAB-PUB-08-006-A-CD}

\title[Optimal parameter estimation for thermal noise]{Optimal estimation of several linear parameters in the presence of Lorentzian thermal noise}

\author{Jason H. Steffen$^1$, Michael W. Moore$^2$ and Paul E. Boynton$^2$}
\affiliation{$^1$ Fermilab Center for Particle Astrophysics M.S. 127 P.O. Box 500 Batavia, IL 60510}
\affiliation{$^2$ University of Washington, Department of Physics, Box 351560, Seattle, WA 98195-1560}
\email{jsteffen AT fnal.gov}

\begin{abstract}
In a previous article we developed an approach to the optimal (minimum variance, unbiased) statistical estimation technique for the equilibrium displacement of a damped, harmonic oscillator in the presence of thermal noise.  Here, we expand that work to include the optimal estimation of several linear parameters from a continuous time series.  We show that working in the basis of the thermal driving force both simplifies the calculations and provides additional insight to why various approximate (not optimal) estimation techniques perform as they do.  To illustrate this point, we compare the variance in the optimal estimator that we derive for thermal noise with those of two approximate methods which, like the optimal estimator, suppress the contribution to the variance that would come from the irrelevant, resonant motion of the oscillator.  We discuss how these methods fare when the dominant noise process is either white displacement noise or noise with power spectral density that is inversely proportional to the frequency ($1/f$ noise).  We also construct, in the basis of the driving force, an estimator that performs well for a mixture of white noise and thermal noise.  To find the optimal multi-parameter estimators for thermal noise, we derive and illustrate a generalization of traditional matrix methods for parameter estimation that can accommodate continuous data.  We discuss how this approach may help refine the design of experiments as they allow an exact, quantitative comparison of the precision of estimated parameters under various data acquisition and data analysis strategies.
\end{abstract}

%\submitto{\CQG}

\maketitle

\section{Introduction}

%After more than two centuries, the principle of the torsion pendulum still underlies the design of instruments to measure ultra-small electrical or gravitational torques.  Recent advances in the performance of these devices depend primarily on improvements in electronic readout systems and the care taken to make optimal or near optimal statistical inference considerations the basis for design of both hardware and measurement protocol \citep{boynton00,hoyle04}.  The current state of this art is driven by inevitable demands for experimental precision to approach the fundamental limit posed by thermal noise as expressed by the fluctuation-dissipation theorem.  In a previous article we developed an approach to the optimal (minimum variance, unbiased) statistical estimation technique for the equilibrium displacement of a damped, harmonic oscillator in the presence of thermal noise (\citet{moore05}, hereafter Paper~I).

%While that work introduces a useful theoretical treatment and some interesting conclusions, its scope is quite limited because it addresses only the case of a single, linear parameter.  For a torsion pendulum under the influence of an external torque this single parameter fails to distinguish between the equilibrium displacement of the pendulum mass and its equilibrium displacement in the absence of the external torque.  Moreover, a real torsion fibre under load slowly unwinds, also requiring parameterization.  Consequently, even without considering pendulum oscillations, a multi-parameter model must be employed.

In spite of the passage of more than two centuries, the principle of the torsion pendulum still underlies the design of instruments to measure ultra-small electrical or gravitational torques.  Ever-improving techniques for suppressing sources of systematic error in these experiments have motivated corresponding reductions in statistical measurement error.  Parallel improvements in optical design and low-noise electronics have previously met this challenge, but statistical uncertainty is inevitably approaching the fundamental thermal-noise limit posed by the fluctuation-dissipation theorem.  Further improvement in signal-to-noise ratio for ambient temperature experiments may be possible only by making optimal or nearly optimal statistical inference considerations a basis for the design of measurement protocol, data analysis, and even instrumentation hardware.

In a previous article we laid the groundwork for an approach to the optimal (minimum variance, unbiased) statistical estimation technique for the parameters of a damped, harmonic oscillator in the presence of thermal noise (\citet{moore05}, hereafter Paper~I), having found no readily accessible treatment of this problem in the literature.  While that work introduces the foundation of a useful methodology, its scope is quite limited because it addresses only the case of a single linear parameter, the displacement of pendulum orientation.  For a torsion pendulum under the influence of an external torque, a simple displacement parameter does not distinguish between the equilibrium displacement of a pendulum mass in the presence versus absence of an external torque, the signature of a torque ``signal''.  Moreover, a real torsion fibre under load slowly unwinds, also requiring parameterization.  Consequently, even without including pendulum oscillations, a multi-parameter model must be employed.

The purpose of this article is to extend the methodology presented in Paper~I to include the optimal estimation of several linear parameters.  As in the case of a single linear parameter, we note that traditional methodologies for estimating multiple parameters fail for a high-$Q$, long-period oscillator subject to thermodynamic fluctuations.  The problem faced stems from the oscillator's response to the thermal bath; the environment drives the oscillator with equal power at all frequencies and the oscillator response is a Lorentzian which peaks sharply at the resonant frequency.  Just as in the single parameter case, in order to minimize the variance in the multi-parameter estimators, one must suppress the variance contribution from this resonance peak.

Superficially, the issues presented by the resonance peak, the recovery of a signal, and the unwinding behaviour of the torsion fibre, appear straightforward.  One might consider treating the noise as white, then address each issue serially to produce a composite measurement approach.  First, to remove the effects of the resonance one could use data from an integer number of oscillation periods as illustrated in Paper~I.  Second, to break the degeneracy between the free orientation of the pendulum mass and its orientation in the presence of a source mass, one could modulate the signal.  Then by choosing the modulation frequency to be distinct from, but commensurate with the oscillation frequency, and also choosing the duration of the data sample to be an integer number of both oscillation and signal periods, the orthogonality of the fitted parameters over the measurement interval could be maintained.

Finally, in order to account for the secular unwinding behaviour of the torsion fibre one could include in the model to which the data are fit a few low-order polynomials.  However, these polynomials are generally not orthogonal to the free oscillation of the pendulum.  Consequently, the estimating function for the signal is no longer orthogonal to the resonance peak.  At this point it becomes clear that such a casual approach must give way to a proper multiple-parameter fitting scheme that filters the contribution from the resonance peak.  As already mentioned, the absence from the literature of explicit techniques for optimal parameter estimation in the analysis of high-$Q$ oscillators in experiments whose precision may approach the limitation posed by thermal noise has motivated our search for a practical solution to this problem.

To set a context for presenting the optimal solution, we discuss two, straightforward, approximate methods that suppress the contribution of the resonance peak to the variances of important model parameters.  The first, which we call the trigonometric method, is to fit the data optimally for white noise while including the sine and cosine oscillation amplitude parameters of the pendulum.  Doing so ensures that the estimators for the remaining parameters---particularly those that correspond to relevant observables---are immune to the variance contribution from the resonance.  

The second method, which we call the \ew\ method, is to pre-filter the data by adding a given realization of pendulum motion to itself displaced by half an oscillation period, thereby eliminating the effects of resonant oscillations to a large extent.  These pre-filtered data are then fit optimally for white noise.  The \ew\ method was employed by the E\"ot-Wash group at the University of Washington in their early torsion pendulum experiments (e.g. \citet{su94}).  While that group has refined this method extensively \citep{kapner07}, we selected the \ew\ estimation technique since it provides a simple, contrasting approach to the trigonometric estimator.  We analyze these two methods as examples of how we can understand and compare their behaviours relative to the optimal estimate derived using the parameter estimation formalism introduced in Paper~I and extended here.  Our purpose in this work is to illustrate the application of this formalism and to point out its advantages, not to critique the methods employed by others.

While these two estimators generally have unequal parameter variances due to their different weighting of the data, both methods remove much of the contribution of the resonance peak to the variances because the associated filter functions, when expressed in the Fourier representation, have a notch with a quadratic minimum at the resonant frequency.  They also share the advantage that variances remain finite when white displacement noise is superposed on the thermal noise as is the case in any real experiment.  By comparison, the variance of the optimal thermal-noise estimators diverge when white noise is present because their displacement estimating functions include Dirac delta functions (see Paper~I).  The ubiquity of white noise motivates us to include these approximate methods in our discussion because they represent the schemes that experimentalists actually employ, and likely will continue to employ in the future.

Since the optimal estimation method is pathological in the presence of white noise, comparing it with approximate methods may appear little more than an interesting exercise because an experiment can never be conducted in this idealized context.  However, such results establish a firm lower bound to the variance of a parameter estimator enabling one to weigh the benefit of improving the experimental methods to lower the uncertainty in a measurement against the labor necessary to implement such modifications.

Moreover, asking which approximate method is superior in a real experiment brings up matters of judgment.  An experimentalist must choose an analysis method and justify the choice in light of many aspects of the experiment at hand.  Design questions regarding the advantages of a particular length of a data sample, the most appropriate phase of the modulated signal, or the effect of adding additional parameters to the model can be answered with confidence for the case of white noise.  For thermal noise, however, the answers to such questions appear unresolved in the literature.  In this paper we set forth a basis for such choices.

The major premise of Paper~I is that transforming to the driving-force basis---where the force noise is clearly white---allows one to readily determine the optimal thermal noise filter for the purpose of estimating a single parameter from a sample of continuous data.  Here we extend this idea to multi-parameter models.  In addition, we demonstrate a second point that was not emphasized in Paper~I: this white-noise driving-force basis is more natural than the displacement basis for developing, analyzing, and comparing different parameter estimation methods---one's intuition is better served in this basis.  Given the insight gained by working in the driving-force basis and the tools we develop to fit multi-parameter models to continuous data, questions about the choices mentioned above can be readily answered with confidence and precision.

We begin our discussion with a brief review of the notation used in Paper~I.  Then we present a comparison of the single-period, optimal thermal-noise estimator of the equilibrium displacement (derived in Paper~I) with a single-period boxcar estimator because the latter, a uniform average, is the foundation of the two approximate estimation methods we consider.  We show in Section \ref{transformsection} how to transform an estimating function from its displacement-basis representation into the basis of the driving force---the inverse of the transformation presented in Paper~I.  We continue with a comparison of optimal, trigonometric, and \ew\ single-parameter estimators of the equilibrium displacement for various durations of the data sample, and then present a method to construct (in the driving-force basis) an estimator that performs well in the case of a mixture of white and thermal noise.  The derivation of this new estimator further demonstrates the advantage of working in the basis of the driving force.

Following this derivation is a digression to develop the methods required to extend the optimal single-parameter, thermal-noise estimators to the multi-parameter case.  The basic result is that optimal multi-parameter estimating functions are orthonormal linear combinations of the optimal single-parameter estimating functions derived in Paper~I.  We are then ready to compare the full, multi-parameter versions of the optimal, trigonometric, and \ew\ methods for a modulated signal while also fitting constant and linear polynomial terms to accomodate fibre unwinding.  Finally, we consider a different thermal noise model that is of particular interest for torsion pendulum experiments.  In moderately high vacuum torsion pendulum experiments, thermal noise results primarily from torsion fibre internal loss mechanisms.  Empirically, the driving-force power spectral density in this case is better approximated by a $1/f$ spectrum rather than the white spectrum of classical dashpot resistance~\cite{saulson90}.  Although the optimal methods we develop no longer strictly apply, we investigate how the thermal-noise optimal, trigonometric, and \ew\ variances compare in the presence of $1/f$ noise; this in addition to similar comparisons for dashpot thermal noise, white noise, and mixed white and thermal noise.

We make two preliminary comments on the presentation of and approximations used in this work.  First, in order to inform the reader's experimental intuition, we choose to present concrete examples based on relevant choices for sample duration, signal frequency, etc.  These examples and the estimation methods that we analyze were selected solely to provide illustrations of the formalism that we develop here and in Paper~I.  Second, because the quality factor, $Q$, is typically several thousand in today's torsion pendulum experiments, we simplify our calculations by truncating results to leading order in $1/Q$.  As mentioned above, traditional methods of parameter inference fail for a high-$Q$, long period oscillator---precisely the regime where this truncation is justified.  Making this approximation here, however, differs from Paper~I in which some results were presented as exact, closed-form expressions.  Note that this work does not present the analysis of any particular experiment.  Rather, it presents a method by which the performance of different parameter estimators can be compared under various noise environments and it demonstrates the conceptual advantages of developing parameter estimators in the basis of the thermal driving force.

\section{Optimal Estimator vs. Boxcar Estimator for a Single Period\label{oneperiod}}

As a point of departure, we first compare two estimators for the equilibrium displacement of the oscillator in the presence of thermal noise, the optimal estimator and the boxcar estimator.  The latter, a simple, uniform average of the time-domain data, is optimal for the case of white displacement noise.

\subsection{Linear Single-Parameter Estimates}

We recall from Paper~I (Eq. 10) that a general parameter estimate for a continuous time series is given by
\begin{equation}
\hat{p} = \fullint e_{\hat{p}}(t) x(t) dt
\end{equation}
where $x(t)$ is the data (for a torsion pendulum $x(t)$ is the measured angular displacement), and the estimating function $e_{\hat{p}}(t)$ is obtained by properly normalizing a filter function.  We again use the convention that a capital letter (e.g. $X(t)$) represents an ensemble of realizations which are represented by lower case letters (e.g. $x(t)$).  The variance of the corresponding parameter estimator, expressed in the Fourier representation (Paper~I, Eq. 14), is
\begin{equation}
\text{var}(\hat{P}) = \frac{1}{2}\fullint F^2\left[ e_{\hat{p}}(t);\nu \right]S\left[\delta X(t);\nu\right]d\nu
\label{varequation}
\end{equation}
where the Fourier energy density (FED) $F^2 [ e_{\hat{p}}(t);\nu ]$ is the square of the Fourier transform of the estimating function and $S[\delta X(t);\nu]$ is the power spectral density (PSD) of the noise ensemble $\delta X(t)$.  Our goal is to calculate the parameter estimator variances to leading order in $1/Q$.  For thermal noise, the PSD already contains a factor of $1/Q$, and so we need express the estimating functions only to zeroth order in $1/Q$ (that is, we may use the estimating functions that would be suitable for an undamped oscillator).

\subsection{Single-Period Optimal Thermal-Noise Estimator\label{singleperiodop}}

In Paper~I we find the optimal estimate of the equilibrium displacement of the oscillator in the presence of thermal noise.  To zero-order in $1/Q$, the single-period, optimal estimate is (c.f. Paper~I, Eqs. 7 \& 44)
\begin{equation}
\hat{c}^{\op} = x_m + \frac{v_f-v_i}{2 \pi \omega_0}
\label{opthermest}
\end{equation}
where $x_m$ is the time-average position of the oscillator, $\omega_0$ is the (undamped) oscillation frequency, and $v_i$ and $v_f$ are the initial and final velocities of the oscillator.  The corresponding estimating function is (c.f. Paper~I, Eqs. 42 \& 43)
\begin{equation}
e_{\hat{c}}^{\op}(t) = \frac{\Theta(t;t_i,t_f)}{\tau_0} + \frac{-\delta'(t-t_f) + \delta'(t-t_i)}{2 \pi \omega_0},
\label{opthermep}
\end{equation}
where $\tau_0$ is the period of the pendulum, $\Theta(t;t_i,t_f)\equiv \theta(t-t_i)-\theta(t-t_f)$ is the boxcar function, and $\delta'(t)$ is the time derivative of the Dirac delta function.  We call this estimate a ``force-only'' estimate because its estimating function is orthogonal to a free-oscillation transient of arbitrary amplitude and phase.  That is,
\begin{equation}\label{req1}
\fullint e_{\hat{c}}^{\op}(t) \cos(\omega_0 t) dt = 0
\end{equation}
and
\begin{equation}\label{req2}
\fullint e_{\hat{c}}^{\op}(t) \sin(\omega_0 t) dt = 0.
\end{equation}

All such force-only estimators have an estimating function that can be expressed in the driving-force basis.  As shown in Paper~I, the optimal displacement estimating function for thermal noise is a boxcar in the force basis (c.f. Paper~I, Eq. 46)
\begin{equation}
y_{\hat{c}}^{\op} = \frac{\Theta(t;t_i,t_f)}{2\pi m \omega_0}
\end{equation}
where $m$ is the mass of the oscillator.  Since thermal noise looks white in the force basis, the variance of the estimator is readily calculated in the time domain,
\begin{equation}
\begin{split}
\text{var}(\hat{C}^{\op}) &= 2k_{\bsub}T\xi \fullint y^2 dt \\
&= \frac{\sigma^2}{\pi Q_0}
\end{split}
\label{varcoptherm}
\end{equation}
where $\sigma^2=k_{\bsub} T/\kappa$ and $k_{\bsub}$ is the Boltzmann constant, $T$ is the absolute temperature of the thermal bath, $\kappa$ is the torsional spring constant, $\xi$ is the damping coefficient, and $Q_0 = m \omega_0/\xi$.

\subsection{Transforming Between Driving-Force Basis and Displacement Basis\label{transformsection}}

In our previous work we showed that an estimating function in the force basis, $y_{\hat{p}}(t)$, can be transformed into the corresponding displacement basis estimating function by using the transpose equation-of-motion operator $\Omega^T$ (c.f. Paper~I, Eqs. 40 \& 41)
\begin{equation}
\begin{split}
e_{\hat{p}}(t) &= \Omega^T\left[y_{\hat{p}}(t)\right]\\
&= \left(m \frac{d^2}{dt^2} -\xi \frac{d}{dt} + \kappa \right) y_{\hat{p}}(t).
\label{eqofmotiontrans}
\end{split}
\end{equation}
For example, in Paper~I $e_{\hat{c}}^{\op}(t)$ was derived from $y_{\hat{c}}^{\op}(t)$ by this method.  As we compare the results of various estimation techniques we wish to exploit the simplifications that follow from working in the force basis where the noise power spectrum is white.  Doing so requires that we obtain the force basis estimating function from the displacement basis estimating function---the inverse of what we have done before.

A straightforward approach to find $y_{\hat{p}}$ from $e_{\hat{p}}$ is to consider the equation of motion relating the driving force to the time series
\begin{equation}
\begin{split}
\mc{F} &= \Omega[x(t)]\\
&= \left( m\frac{d^2}{dt^2}+\xi \frac{d}{dt} + \kappa \right) x(t)
\label{eqofmotion}
\end{split}
\end{equation}
where $\Omega$ is the equation of motion operator.  We know from the solution to this differential equation that, to leading order in $1/Q$,
\begin{equation}
\begin{split}
x(t) =& x_i\cos(\omega_0 (t-t_i))+\frac{v_i}{\omega_0}\sin(\omega_0 (t-t_i))\\
&+\frac{1}{m\omega_0}\int_t^{\infty}\mc{F}(t')\sin(\omega_0 (t' - t))dt'.
\label{odesolution}
\end{split}
\end{equation}
For a high-$Q$ oscillator the quantity $\xi \simeq 0$ is very small and the equation of motion operator is well approximated by its transpose, $\Omega \simeq \Omega^T$.  Substituting $\Omega$ for $\Omega^T$ in equation (\ref{eqofmotiontrans}) one obtains the solution for $y_{\hat{p}}$ that is similar to that in equation (\ref{odesolution}) 
\begin{equation}
y_{\hat{p}}(t) = \frac{1}{m\omega_0}\int_t^{\infty}e_{\hat{p}}(t')\sin(\omega_0 (t' - t))dt'
\label{yfrome}
\end{equation}
except that the boundary value terms vanish because the estimating function must be identically zero outside of the time series.  It is important to note that the $y_{\hat{p}}(t)$ that results from this equation is valid only when $e_{\hat{p}}(t)$ satisfies both (\ref{req1}) and (\ref{req2})---it is force-only, sensitive to the driving force but not the free oscillation of the pendulum.

\subsection{Single Period Boxcar}

Let us now apply this method to a boxcar estimator (unless otherwise noted the term ``boxcar estimator'' will hereafter refer to one constructed in the displacement basis).  We first calculate the force basis estimating function that corresponds to the boxcar estimator to zeroth order in $1/Q$
\begin{equation}
\begin{split}
y_{\hat{c}}^{\boxsub}(t) &= \frac{1}{m\omega_0}\int_t^{\infty} \frac{\Theta(t';-\tau_0/2,\tau_0/2)}{\tau_0} \sin(\omega_0(t'-t)) dt' \\
&= \frac{\Theta(t;-\tau_0/2,\tau_0/2)}{2\pi m\omega_0}\left( 1+\cos(\omega_0 t)\right).
\end{split}
\label{yboxone}
\end{equation}
The variance in the corresponding parameter estimator is now
\begin{equation}
\begin{split}
\text{var}(\hat{C}^{\boxsub}) &= 2k_{\bsub}T\xi \fullint y^2 dt \\
&= \left( \frac{3}{2} \right) \frac{\sigma^2}{\pi Q}\\
&= \frac{3}{2}\text{var}(\hat{C}^{\op}).
\end{split}
\label{varcboxone}
\end{equation}
Thus, the penalty for using the boxcar estimator instead of the optimal estimator is a factor of 3/2 increase in the parameter variance.

To understand the source of this 50\% penalty we consider (\ref{yfrome}) where it follows that a force-only $e_{\hat{p}}(t)$ which is bounded (that is, never infinite) results in a $y_{\hat{p}}(t)$ that smoothly transitions to zero at both $t_i$ and $t_f$---it is zero and has zero slope at the endpoints.  The cosine term in (\ref{yboxone}) serves to match these boundary conditions.  Since the cosine function is orthogonal to a constant function over one period, the amplitude of the boxcar in (\ref{yboxone}) must be the same as in the optimal in order for the estimate to be unbiased.  Finally, because the average value of cosine squared is 1/2, the variance increases by 50\%.  Conversely, if one begins with a force-only $e_{\hat{p}}(t)$ that is unbounded (e.g. has $\delta(t)$'s or $\delta'(t)$'s at the endpoints), then the resulting $y_{\hat{p}}(t)$ has discontinuous jumps that correspond to measurements of the boundary conditions of the oscillator.  The fact that the boxcar estimator does not utilize this boundary information results in its increased uncertainty in the values of the estimated parameters.

\section{Comparison of Optimal, Trigonometric, and Autosum Estimators\label{comparisonsec}}

The factor of 3/2 in (\ref{varcboxone}) leads one to ask if additional data improves this result and if the variance in the boxcar estimator ultimately converges to the optimal value.  It turns out that neither occurs under the approximations under consideration.  Yet, this optimal filter for white displacement noise can be used to construct an estimator that performs much better when thermal noise dominates.  Here we develop and study two estimators, the \ew\ and the trigonometric estimators, both of which are based upon the boxcar estimator.  Following that, we compare the performance of these estimators for the cases of white noise and ``mixed noise'', which has both thermal and white noise components.  Finally, we present the construction of a new estimator that performs well for cases of mixed noise.  The development of this new estimator takes advantage of the intuitive relation between the performance of an estimator and its functional form in the driving-force basis.  This development is facilitated by knowledge of the form of the optimal estimator.

\subsection{Thermal Noise and Multiple Period Estimators\label{secthreea}}

Following the calculations in Section \ref{singleperiodop} one can show that the variance of the optimal estimator times the duration of the data, $\tau$, is
\begin{equation}
\var (\hat{C}^{\op}) \times \tau = \frac{2\sigma^2}{\omega_0 Q_0} = \text{constant}.
\end{equation}
Thus, for three periods of data, the optimal variance has decreased by a factor of 1/3.  Similarly, for three periods of data the variance in the boxcar estimator has decreased by a factor of 1/3.  This pattern continues for data that span an integer multiple of oscillation periods.  To leading order in $1/Q$ the variance in the boxcar estimator remains a factor of 3/2 larger than the optimal.

\begin{figure}
\includegraphics[width=0.45\textwidth]{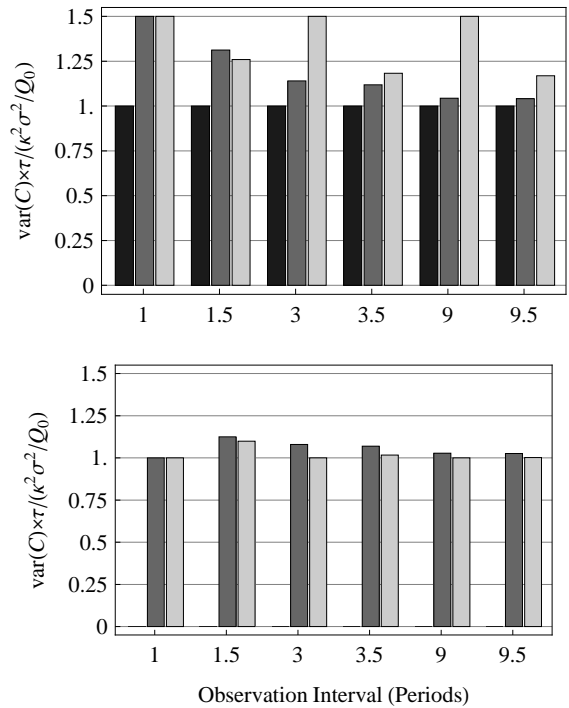}
\caption{Comparison of variances in the optimal estimator (black), \ew\ estimator (dark gray), and the trigonometric estimator (light gray) in the presence of thermal noise (top panel) and white noise (bottom panel) for observation intervals equal to 1, 1.5, 3, 3.5, 9, and 9.5 oscillation periods.  For white noise, the variance in the optimal thermal-noise estimator would be infinite and is therefore not shown.  Here, the white noise PSD is normalized to match the zero frequency limit of the thermal noise PSD [see (\ref{whitenormalize})]}
\label{variancecomp1}
\end{figure}

The \ew\ estimator improves upon this by averaging the time series with itself displaced by 1/2 period.  However, doing so requires an additional 1/2 period of raw data.  The 3.5-period \ew\ estimating function is
\begin{widetext}
\begin{equation}
\begin{split}
e_{\hat{c}}^{\ewsub} =& \frac{\Theta(t;-1.75\tau_0,1.25\tau_0) + \Theta(t;(-1.75+0.5)\tau_0,(1.25+0.5)\tau_0)}{6\tau_0}\\
=& 0.1667\frac{\Theta(t;-1.75\tau_0,-1.25\tau_0)}{\tau_0} + 0.3333\frac{\Theta(t;-1.25\tau_0,1.25\tau_0)}{\tau_0} + 0.1667\frac{\Theta(t;1.25\tau_0,1.75\tau_0)}{\tau_0}
\end{split}
\end{equation}
where the ``as'' superscript represents the ``autosum'' estimating technique.  From (\ref{yfrome}) this gives the force-basis estimating function
\begin{equation}
%\begin{split}
y_{\hat{c}}^{\ewsub} = \frac{0.1667}{2\pi m \omega_0} \bigl( \left(1-\sin(\omega_0 t)\right)\Theta(t;-1.75\tau_0,-1.25\tau_0) + 2 \Theta(t;-1.25\tau_0,1.25\tau_0) + \left(1+\sin(\omega_0 t)\right)\Theta(t;1.25\tau_0,1.75\tau_0)\bigr).
%\end{split}
\label{ewthreeptfive}
\end{equation}
\end{widetext}
The variance in the corresponding estimator is $112\%$ of optimal (this numerical result and others from the following discussion will be represented in Figure \ref{variancecomp1}).

\begin{figure}
\includegraphics[width=0.45\textwidth]{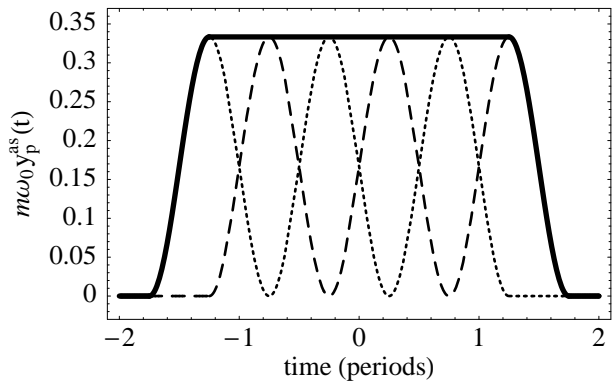}
\caption{Graph of the 3.5 period \ew\ estimating function.  The two dashed curves correspond to the data and the data displaced by half of one oscillation period.  Their sum, the \ew\ estimating function, is the solid curve.  Note that the sinusoidal terms contribute only near the ends of the observation interval, allowing the variance in the equilibrium displacement estimator to approach the optimal value as the duration of the data increases.}
\label{eotwashpic}
\end{figure}

To illustrate why the \ew\ estimator is superior to the boxcar estimator ($112\%$ of optimal compared with $150\%$), recall that the optimal estimating function is a boxcar in the force-basis.  In this basis, the displacement-basis boxcar is a constant plus a sinusoid whereas the \ew\ estimating function is simply a constant with smooth transitions to zero near the endpoints (sinusoidal profile).  Figure \ref{eotwashpic} shows the 3.5 period force-basis representation of the \ew\ estimating function as the sum of its parts.  When viewed in the force basis, the \ew\ estimating function can be seen to be a better approximation to the optimal estimating function than is the displacement-basis boxcar.

Comparing the 3-period boxcar estimator with a 3.5-period \ew\ estimator is not entirely a fair comparison.  The variance of a 3-period \ew\ estimator, which results from the sum of two mutually displaced 2.5-period boxcar estimating functions, is $114\%$ of optimal---still significantly better than the boxcar estimator.  In addition, while the constituent 2.5-period boxcars are not individually orthogonal to the pendulum oscillation the 3-period \ew\ estimator is.  Indeed, any \ew\ estimator is a force-only estimator because the unique summation scheme removes the effect of the free oscillation to zero order in $1/Q$; thus satisfying the conditions (\ref{req1}) and (\ref{req2}).  On the other hand, any boxcar estimator that is a half-integer number of periods in duration will suffer a significant penalty of order $Q$ in variance inflation because of contamination from the pendulum oscillation (as shown in Paper~I).

If we were to use the \ew\ estimator for a single period of data, then the resulting variance in the parameter estimator would be 150\% of the optimal---identical to the variance for the boxcar.  This can be seen the force-basis representation of this estimator shown in Figure \ref{eotwashpic} where, if the flat middle portion of the \ew\ estimating function were to vanish, the remaining function is equivalent to the force-basis representation of the displacement-basis boxcar, a sinusoid.  Thus, within a few periods of data the \ew\ estimator improves dramatically over the boxcar estimator.

The variance in the \ew\ estimator continues to converge toward the optimal estimator as the observation interval increases because the constant portion of the force-basis estimating function constitutes a larger and larger fraction of the total duration.  In the force basis, the primary difference between the \ew\ estimator and the optimal estimator is the smooth transition at the edges from the constant value to zero for the former as opposed to the step function transition for the latter.  For thermal noise, abrupt edges characterize the optimal estimator; however, the consequence of those edges is shown when white noise, which lacks a high-frequency cutoff, is present.  We discuss the effects of these discontinuities in Section \ref{whitenoise}.

We now turn our attention to the trigonometric approach---a different, but straightforward improvement of the boxcar estimator.  The trigonometric estimator results from including the sine and cosine components of the free oscillation of the pendulum in the fit.  This can be accomplished, for example, using a general form of the estimating function
\begin{equation}
e_{\hat{p}}(t) = \Theta(t;t_i,t_f)\left(k_0 + k_1 \cos(\omega_0 t) + k_2 \sin(\omega_0 t)\right)
\end{equation}
and choosing the values of $k_0$, $k_1$, and $k_2$ such that the constraints (\ref{req1}) and (\ref{req2}) are satisfied and that the normalization is correct (as stated in equation (12) of Paper~I).  The trigonometric approach produces reasonable estimators for any duration of the sample.  The 3.5-period estimating functions in the displacement and force bases are then
\begin{equation}
e_{\hat{p}}^{\trsub} = \frac{\Theta(t;-1.75\tau_0,1.75\tau_0)}{\tau_0}(0.2905 + 0.0528\cos(\omega_0t))
\end{equation}
and
\begin{equation}
\begin{split}
y_{\hat{p}}^{\trsub} = & \frac{\Theta(t;-1.75\tau_0,1.75\tau_0)}{2\pi m\omega_0}\\
\times & (0.2905 + 0.0264\cos(\omega_0t) +0.0264\omega_0t\sin(\omega_0t))
\end{split}
\end{equation}
respectively.  The variance of the 3.5-period trigonometric estimator is 118\% of optimal, much better than 150\% for the 3-period trigonometric estimator (which is equivalent to the 3-period boxcar).  Not only is this estimator acceptable, but in terms of variance (not variance times time) the 3.5-period trigonometric estimator is better than the 4-period trigonometric (or boxcar) estimator.

\begin{figure}
\includegraphics[width=0.45\textwidth]{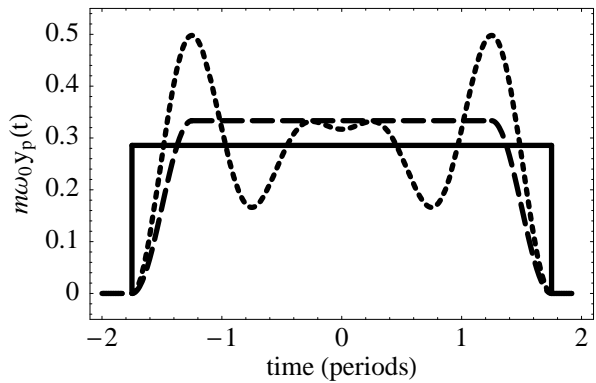}
\caption{A comparison of the 3.5-period optimal (solid), \ew\ (dashed), and trigonometric (dotted) force-basis estimating functions.}
\label{comparison}
\end{figure}

Although its force-basis estimating function always contains oscillatory terms and does not, therefore, resemble the constant profile of the optimal estimating function, the half-integer period trigonometric estimator is better than its integer period counterpart because there is a triangular envelope on the sine component of the oscillation which suppresses the contribution from that term.  Figure \ref{comparison} overlays the 3.5-period optimal, \ew , and trigonometric estimating functions.  For longer observation times, the trigonometric estimator remains at 150\% of optimal for integer periods and converges to $7/6 = 117\%$ of optimal for half-integer periods, if examined to first order in $1/Q$---this is unlike the \ew\ estimator which converges identically to the optimal.  The top panel of Figure \ref{variancecomp1} compares the variances in these three estimators multiplied by the duration of the sample for 1, 1.5, 3, 3.5, 9, and 9.5 periods of data.  The motivation for using 3 and 9 periods stems from our examination (later in the discussion) of modulated force signals.  The corresponding half-integer periods are included because the \ew\ estimator uses raw data that span a half-integer number of periods.

%We take the values of 9 and 9.5 periods, as selected by the \ew\ group, to be representative of practice.  Our original choice of examining 3 and 3.5 periods constitute useful samples of intermediate duration, and matches a convention of the \ew\ group to modulate their force at 2/3 of the resonant frequency\citep{gundlach97}.  Three periods of data also correspond to the first choice for which both an integer number of modulation periods and an integer number of pendulum oscillation periods occur.  The 3.5-period results are shown because the \ew\ estimator uses raw data that spans a half-integer number of periods.

Since the variance in the \ew\ estimator approaches that in the optimal estimator as the duration of the data increases, a practical limit to the length of the observations is set primarily by factors that lie beyond mathematical considerations.  These could include unavoidable non-Gaussian disturbances that partition the realization or low-frequency noise (drift).

\subsection{White Noise and Mixed Noise\label{whitenoise}}

So far we have examined the effects of thermal noise alone on the variances in these estimators.  A real system will always display some white displacement noise as well.  For white noise the variance of a given parameter estimator is most easily obtained in the displacement basis
\begin{equation}
\text{var}(\hat{P}_{\wh}) = \frac{\eta}{2}\fullint e_{\hat{p}}^2 dt
\end{equation}
where $\eta$ is the (constant) PSD of the noise.  Since the displacement-basis estimating functions involve derivatives of the force-basis estimating functions (see (\ref{eqofmotiontrans})), only those force-basis estimating functions that are smooth (i.e. second-order differentiable) will have a well defined variance in the presence of white displacement noise.  The step-function transition exhibited by the estimating functions that are optimal for thermal noise will have Dirac delta functions and their derivatives in the displacement basis.  Thus, the high-frequency components of white noise are infinitely amplified by those delta functions and the variances of the estimators diverge to infinity.

The lower panel of Figure \ref{variancecomp1} displays the variances in each of the two approximate estimators for the case of white noise.  For comparison, we normalize the white noise PSD so that it is equal to the zero frequency limit of the thermal noise PSD.  That is,
\begin{equation}
\begin{split}
S[\delta X_{\wh}(t);\nu] &= S[\delta X_{\thsub}(t);\nu=0]\\
&= \frac{4k_{\bsub}T\xi}{\kappa^2}\\
&= \frac{4\sigma^2}{Q_0\omega_0}.
\label{whitenormalize}
\end{split}
\end{equation}
We see from Figure \ref{variancecomp1} that the trigonometric approach is always superior to the \ew\ approach for the case of white noise, although the difference may be small.

The optimal estimator for white displacement noise is the boxcar estimator.  Both the trigonometric and the \ew\ estimators are equivalent to the boxcar estimator for a single period of data, as can be seen in both panels of Figure \ref{variancecomp1}.  For 1.5 periods, the trigonometric estimator has smaller variance than the \ew\ estimator for both noise processes.

From the calculations in this section we can do more than qualitatively compare the two approximate estimators; we can identify the ratio of white to thermal noise where one approach becomes superior to the other.  Following the PSD normalization in (\ref{whitenormalize}), we state the mixing ratios as the fraction of the total noise power contributed from a given noise process, white or thermal.  Table \ref{noisemixresults} outlines the results of this test.

\begin{table}
\caption{\label{noisemixresults}Noise mixing ratios for various sample durations where the trigonometric estimators become superior to the \ew\ estimators.  The \ew\ estimators are superior when the white noise contribution is less than the stated amount.}
\begin{center}
\item[]
\begin{tabular}{lcccc}
\hline
Periods              & 3    & 3.5  & 9    & 9.5  \\
\hline \hline
White Noise Fraction & 82\% & 55\% & 94\% & 84\% \\
\hline
\end{tabular}
\end{center}
\end{table}

These results indicate that a knowledge of the dominant noise components and their relative importance is a necessary ingredient informing the choice of the data acquisition protocol, sample duration, and the parameter estimation technique to be applied.  Moreover, an intuitive understanding of how the general shape of an estimating function depends upon sample duration---and consequently affects the estimator variance (see Figures \ref{eotwashpic} and \ref{comparison})---may prove a valuable guide when confronted with realistic noise backgrounds or further modifications to the noise PSD.  As an example, we later explore the case for thermal noise with a $1/f$ PSD where we apply the ideas discussed here.

\subsection{Improved Estimator for Mixed Noise}

In this section, we present a method to identify an improved estimator for the case of mixed white and thermal noise.  The development of this new estimator is accomplished in the basis of the driving force which we claim enables a more intuitive understanding of the effects of noise in the system under study.  We will call this estimator the ``\mix '' estimator.

Consider Figure \ref{comparison}, particularly the curves for the \ew\ and the optimal thermal estimating functions.  The sinusoidal transition near the ends of the \ew\ estimating function is the primary difference between that estimator and the optimal thermal estimator.  These transitions are responsible for the excess variance in the \ew\ estimator in the presence of thermal noise.  Yet, when white noise is present, it is these smooth transitions that prevent the variance in the \ew\ estimator from diverging to infinity.  In the case where white noise makes a very small contribution to the total noise, neither the optimal thermal-noise estimator (which fails completely) nor the \ew\ estimator is the best choice.  What is required is an estimating function that resembles more closely the optimal estimating function but that has smooth, rather than abrupt, transitions near the endpoints.

One solution is to have a flat estimating function with sinusoidal transitions near the ends---like the \ew\ estimator---but where the transition frequency can be much faster than the oscillation frequency.  One can define a class of functions, with a single parameter, that are flat in the force basis and that have smooth, sinusoidal transitions at the ends.  This parameter, which determines the steepness of the transition, can be interpreted as the ratio of the transition frequency to the oscillation frequency.  The \ew\ estimator is the special case where the value of the parameter is unity and the transition frequency is equal to the oscillation frequency.

\begin{figure}
\includegraphics[width=0.45\textwidth]{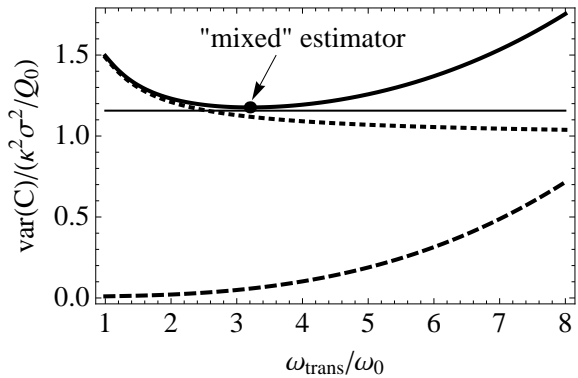}
\caption{This graph shows the estimator variance (solid curve) for the described class of estimators as a function of the transition frequency when the noise is a mixture of 99\% thermal noise and 1\% white noise.  The thin solid line is the minimum variance for this mixture as calculated numerically.  The dotted curve and the dashed curve show the variance contributions from the thermal noise component and the white noise component respectively.  The minimum occurs near 3.2 times the resonance frequency.}
\label{varvscc}
\end{figure}

Now, given a specific mixture of white and thermal noise, one can choose the value of this parameter that minimizes the total variance in the estimators of the oscillator parameters; the result gives the ``\mix '' estimator.  We note that the value of the transition frequency cannot be left completely unconstrained---it must be larger than one over twice the data duration ($\omega_{\text{trans}} \geq 1/2\tau$).  Figure \ref{varvscc} shows the variance in this class of estimators as a function of the ratio of the transition frequency to the oscillation frequency for a single period of data and where a mixture of 99\% thermal noise and 1\% white noise is present.  In this case, the minimum variance (and hence the \mix\ estimator) occurs when the transition frequency is roughly 3.2 times the frequency of the pendulum's free oscillation.

\begin{figure}
\includegraphics[width=0.45\textwidth]{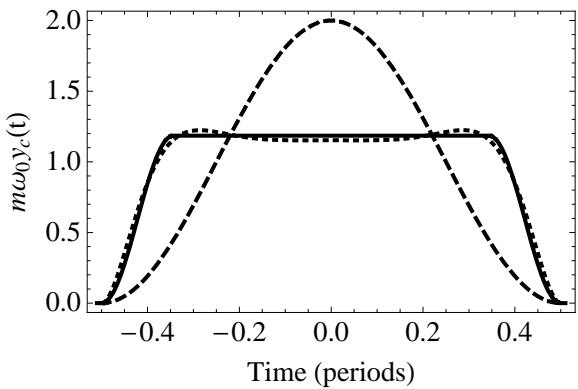}
\caption{Driving-force basis estimating functions for a single period of data where 99\% of the noise is thermal noise and 1\% is white.  The dotted curve is the optimal estimating function, found numerically from 100 discrete points in time.  The dashed line is the boxcar estimating function (equivalent to the single period \ew\ and trigonometric estimating functions).  The solid curve is the \mix\ estimating function that results from the procedure discussed in this section.}
\label{newestimatoryp}
\end{figure}

\begin{figure}
\includegraphics[width=0.45\textwidth]{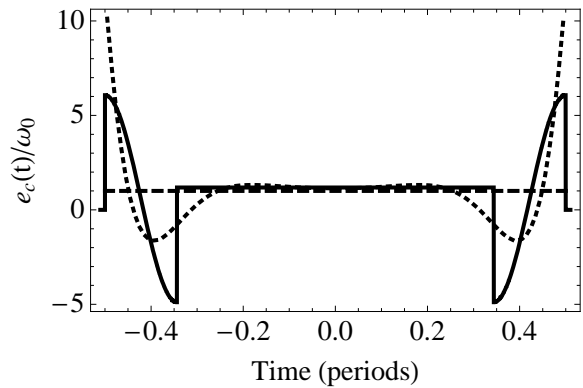}
\caption{Displacement basis estimating functions for a single period of data where 99\% of the noise is thermal noise and 1\% is white.  The dotted curve is the optimal estimating function, found numerically from 100 discrete points in time.  The dashed line is the boxcar estimating function (equivalent to the single period \ew\ and trigonometric estimating functions).  The solid curve is the \mix\ estimating function that results from the procedure discussed in this section.}
\label{newestimatorep}
\end{figure}

Figures \ref{newestimatoryp} and \ref{newestimatorep} show the \mix\ estimating function in the force basis and the displacement basis respectively for the case in question.  These figures also show the respective boxcar estimating functions (equivalent to both the trigonometric and the \ew\ estimating functions) as well as the estimating function that results from a discrete least squares analysis using 100 sample points.

These figures stand as evidence to support our claim regarding the improved intuition that comes from working in the driving-force basis.  Consider Figure \ref{newestimatorep}.  It is not clear that the sinusoidal ``teeth'' at the ends of the displacement-basis representation of the \mix\ estimator are a reasonable approximation to the smoother transitions of the mixed-noise, optimal estimator (and ultimately the derivatives of Dirac delta functions given in Equation (\ref{opthermep})).  If one were working in the displacement basis alone, one must consider the subtle interplay between the dynamical characteristics of the system (e.g. the pendulum mass and torsional spring constant) and how the effects of force noise are manifest in that system before the \mix\ estimator might appear to qualify as an approximation to the optimal version.

On the other hand, from Figure \ref{newestimatoryp} (the representation of the \mix\ estimator in the force basis) it can be seen that the \mix\ estimator is a function that closely resembles the optimal estimator.  Thus, from this perspective, the quantitative agreement shown in Figure \ref{varvscc} (the minimum of the thick solid curve compared to the thin solid line) is not surprising.  Moreover, the \mix\ estimator was constructed in a straightforward manner by taking the \ew\ estimator and allowing the middle portion to become more flattened as the transitions become more steep---approaching the step function transition of the optimal thermal-noise estimator.  Making a similar construction in the displacement basis is less obvious since one must devise an appropriate approximation to a pathological feature (i.e. the derivative of a Dirac delta function) rather than a step function.

The \mix\ estimator is useful for several reasons, of which we mention two.  First, one now has an analytic formulation of an estimator that is optimized for any specific mixture of white and thermal noise.  While the \mix\ estimator may not match the true optimal estimator exactly, it will be the optimal estimator subject to the constraint that it has the described functional form.  The \mix\ estimator will perform at least as well as the \ew\ estimator in the presence of white noise, as well as the optimal thermal estimator in the presence of thermal noise, and it will perform better than either the optimal thermal or the \ew\ estimators for an admixture of white and thermal noise.  As shown in Figure \ref{varvscc} the difference in variance between the mixed-noise optimal estimator (found numerically) and the \mix\ estimator is quite small.

A second use of the \mix\ estimator is that, unlike the \ew\ estimator, this estimator is defined for data durations that are less than 1/2 of an oscillation period in length.  The \ew\ estimator is identical to the boxcar estimator for data that is one period in length.  As the duration of the data approaches 1/2 of an oscillation period, the \ew\ estimator approaches two dirac delta functions located at $\pm 1/4 \tau_0$ and the variance in the \ew\ estimator diverges to infinity.  Since the \mix\ estimator is not constrained to transition at the oscillation frequency, it can be used if there were a need to estimate the parameters of the system using data from only a fraction of the period.

\section{Fitting for Multiple Parameters\label{multiparams}}

Before expanding the material that we have developed thus far to include multiple parameters, we first give a brief review of linear, optimal, multi-parameter fits for discrete data.  We then generalize the discrete case for use in fitting a continuous time series.  For white noise the techniques for discrete data are readily extended to continuous data because the data covariance matrix is a multiple of the identity matrix and thus can be trivially inverted.  For thermal noise, the generalization from discrete to continuous requires that we develop an equivalent formulation of the discrete case that relies on knowledge of the optimal single-parameter filters, which we derive in Paper~I, rather than an explicit inversion of the data covariance matrix.

\subsection{Linear Multi-Parameter Fits for Discrete Data}

For a multi-parameter fit with discrete data, the data from a particular realization are assembled into a data vector and the parameters into a parameter vector
\begin{equation}
\begin{split}
\B{x} & = \{ x_1,x_2, \dots, x_n \} \\
\B{p} & = \{ p_1,p_2, \dots, p_m \}.
\end{split}
\end{equation}
Following the convention of Paper~I, the true, physical value of the parameters will be specified by the vector $\bsym{\rho}$.  The partial derivatives of the data with respect to each of the parameters constitutes the design matrix, $\matx{q}$. To avoid confusion between vectors and matrices, matrices will be indicated by an overbar.  The data vector can then be decomposed into a signal vector, the product of the design matrix with the physical parameter vector, and a noise vector $\delta \B{x}$ which is a realization of the random vector $\delta \B{X}$
\begin{equation}
\B{x} = \matx{q} \ \bsym{\rho} + \delta \B{x}.
\end{equation}
The noise covariance matrix
\begin{equation}
\matx{m}_X = \langle
\delta \B{X}\otimes \delta\B{X}
\rangle.
\end{equation}
characterizes the second moments of the noise ensemble, and a parameter estimate vector is obtained by contracting an estimating matrix $\matx{e}_{\hat{p}}$ with the data vector,
\begin{equation}
\hat{\B{p}} = \matx{e}_{\hat{p}} \ \B{x}.
\end{equation}

Many books (see e.g. \citet{hamilton64}) develop the general theory of optimal, least-squares, parameter estimation for discrete data.  Here we simply quote one of the main results.  The optimal estimating matrix that returns the minimum variance estimate of all parameters in the presence of the noise $\delta \B{X}$ is
\begin{equation}\label{opes}
\matx{e}_{\hat{p}}^{\op} = \left( \matx{q}^T \ \matx{m}_X^{\ -1} \ \matx{q} \right)^{-1} \ \matx{q}^T \ \matx{m}_X^{\ -1}
\end{equation}
We also note that the design matrix and the optimal estimating matrix are pseudo-inverses (a subclass of Moore-Penrose pseudo-inverses) having the property
\begin{equation}
\matx{e}_{\hat{p}}^{\op} \ \matx{q} = \left( \matx{q}^T \ \matx{m}_X^{\ -1} \ \matx{q} \right) ^{-1} \ \matx{q}^T \ \matx{m}_X^{\ -1} \ \matx{q} = \B{I}.
\end{equation}
If this orthonormality condition were not met, a bias would be present in the parameter estimates.

\subsection{Optimal White Noise Multi-Parameter Fit for Continuous Data}

Although the \ew\ method pre-filters the data and the trigonometric method adds two additional parameters, both methods ultimately fit the data in the manner that is optimal for white noise.  Thus it is instructive to consider optimal, white noise, multi-parameter fitting in the continuous limit.  For discrete white noise fits, the data covariance matrix is a multiple of the identity matrix, and the optimal multi-parameter estimating matrix in (\ref{opes}) simplifies to 
\begin{equation}
\matx{e}_{\hat{p}}^{\op} = \left( \matx{q}^T \ \matx{q} \right)^{-1} \ \matx{q}^T \ .
\end{equation}
The transition from discrete to continuous data is then straightforward---the design matrix becomes a vector of functions that is multiplied by the parameter vector to produce the multi-parameter signal.  Thus,
\begin{equation}
x(t;\B{p}) = \B{q}(t) \cdot \B{p}.
\end{equation}
The estimating matrix becomes a vector of functions such that
\begin{equation}\label{paramvec}
\hat{\B{p}} = \fullint \B{e}_{\hat{p}}(t) \ x(t) dt.
\end{equation}
The design vector and the optimal estimating vector of functions satisfy the orthonormality condition,
\begin{equation}
\fullint 
\B{e}_{\hat{p}}^{\op}(t) \otimes \B{q}(t)%} 
dt = \B{I}
\end{equation}
where $\otimes$ represents an outer product.

In analogy with the discrete white noise case, the optimal estimating functions are obtained from the vector of design functions,
\begin{equation}
\B{e}_{\hat{p}}^{\op}(t) = \left( \fullint
\B{q}(t) \otimes \B{q}(t)
dt \right)^{-1} \ \B{q}(t)
\label{whiteeop}
\end{equation}
This equation essentially states that the optimal white noise estimating functions are linear combinations of the single-parameter matched filters such that the estimating functions are orthogonal to the design functions of the remaining parameters\citep{hamilton64}.

\subsection{Optimal Thermal-Noise Multi-Parameter Fit for Continuous Data}

We now consider an alternative formulation of the discrete fit that does not use an explicit inversion of the data covariance matrix.  This formulation is desirable when analyzing continuous data, since inverting the covariance matrix would require finding the Green's function that diagonalizes the covariance operator.  Avoiding this step simplifies the calculations.

Consider the case of a single-parameter fit to discrete data.  The design matrix and estimating matrix are then vectors.  The optimal estimating matrix
\begin{equation}
\B{e}_{\hat{p}}^{\op} = \left( \B{q}^T \ \matx{m}_X^{\ -1} \ \B{q} \right)^{-1} \ \B{q}^T \ \matx{m}_X^{\ -1}
\end{equation}
is also a vector and the term $\left( \B{q}^T \ \matx{m}_X^{\ -1} \ \B{q} \right)^{-1}$ is a multiplicative constant.  From this result, we define the optimal single-parameter filter,
\begin{equation}
\B{f}_{\hat{p}}^{\op} = \B{q}^T \ \matx{m}_X^{\ -1}
\end{equation}
and allow the constant mentioned before to assume the role of normalization.

To extend this to multi-parameter fits, let us provisionally define an optimal multi-parameter filter matrix as
\begin{equation} \label{multifiltervec}
\matx{f}_{\hat{p}}^{\op} \equiv \matx{q}^T \ \matx{m}_X^{\ -1}
\end{equation}
where each row of the filter matrix is the optimal single-parameter filter for the corresponding parameter.  In addition, let us assume that the optimal estimating matrix can be obtained from this filter matrix through multiplication by some square matrix $\matx{n}$.  Since the optimal estimating matrix and the design matrix are pseudo-inverses
\begin{equation}\label{estimatingmatrix}
\matx{e}_{\hat{p}}^{\op} \ \matx{q} = \matx{n} \ \matx{f}_{\hat{p}}^{\op} \ \matx{q} = \B{I}
\end{equation}
the matrix $\matx{n}$ must have the value,
\begin{equation}
\matx{n} = \left( \matx{f}_{\hat{p}}^{\op} \ \matx{q} \right)^{-1}.
\end{equation}
Examination of (\ref{estimatingmatrix}) shows that the estimating matrix is
\begin{equation}
\matx{e}_{\hat{p}}^{\op} = \matx{n} \ \matx{f}_{\hat{p}}^{\op} = \left( \matx{q}^T \ \matx{m}_X^{\ -1} \ \matx{q} \right)^{-1} \ \matx{q}^T \ \matx{m}_X^{\ -1},
\end{equation}
in agreement with the optimal estimating matrix (\ref{opes}).  Thus, our assumption that $\matx{e}_{\hat{p}}^{\op} = \matx{n} \ \matx{f}_{\hat{p}}^{\op}$ is seen to be justified.  Moreover, we now have an equation for the optimal estimating matrix
\begin{equation} \label{opem}
\matx{e}_{\hat{p}}^{\op} = \left( \matx{f}_{\hat{p}}^{\op} \ \matx{q} \right)^{-1} \ \matx{f}_{\hat{p}}^{\op},
\end{equation}
which does not require explicit knowledge of $\matx{m}_X^{\ -1}$.

With equation (\ref{opem}) one can construct the optimal estimating matrix given the optimal single-parameter filter functions regardless of the means by which those functions were identified (recall that each row of the optimal filter matrix is merely the optimal filter obtained from a single parameter fit).  While the expressions for the optimal filter matrix (\ref{multifiltervec}) and the optimal estimating matrix (\ref{estimatingmatrix}) are not obviously generalized for continuous data, taking the expression given in (\ref{opem}) to the continuous limit is more straightforward since the single-parameter filters readily generalize to functions of time.  This generalization was an important part of Paper~I; and for thermal noise, the continuous single-parameter filters can be calculated using equation (40) of that paper.

For continuous data, the resulting vector of optimal estimating functions is obtained from the vector of single-parameter filters and the vector of design functions.
\begin{equation}
\B{e}_{\hat{p}}^{\op}(t) = \left( \fullint
\B{f}_{\hat{p}}^{\op}(t) \otimes \B{q}(t)
dt \right)^{-1} \ \B{f}_{\hat{p}}^{\op}(t)
\end{equation}
Similar to the white noise case, this demonstrates the important result that the optimal thermal-noise estimating functions are linear combinations of the optimal, single-parameter filters.  We apply this information to a multi-parameter model and study its consequences in the next section.

\section{Fitting to a Four-Parameter Model\label{section5}}

The results from this article as well as from our previous work are valid for a constant-force signal.  As stated in the introduction, such a signal is indistinguishable from the normal equilibrium displacement of the oscillator.  To break this degeneracy one must modulate the force signal.  We maintained in Paper~I that our constant-force results were readily applicable to modulated signals; we now show when that assertion is justified.

For the purposes of discussion, we choose to modulate the force signal at 2/3 the oscillation frequency $\omega_s = 2 \omega_0 / 3$.  For this choice of $\omega_s$, the displacement amplitude is a factor of
\begin{equation}
\frac{1}{1-(\omega_s/\omega_0)^2} = \frac{9}{5}
\end{equation}
larger than for a stationary force of the same magnitude.  This raises the question of whether we should compare the constant force results to a modulated signal of the same displacement amplitude or the same driving-force amplitude.  The answer to this question depends upon the experimental approach that one employs for a given experiment.  For a torsion pendulum experiment that modulates the signal by smoothly changing the position of a source mass, it is the magnitude of the applied force that remains constant as the modulation frequency increases.  Thus, a comparison of equivalent driving-force amplitude provides a more useful comparison of signal-to-noise ratio.

We parameterize the oscillator's response as
\begin{equation}
x(t) = \frac{9}{5}\frac{a_s}{\kappa}\cos(\omega_s t) + \frac{9}{5}\frac{b_s}{\kappa}\sin(\omega_s t) + c_0 + c_1t
\label{parameterizationeq}
\end{equation}
so the variance in signal parameters $a_s$ and $b_s$ reflects the same uncertainty in the applied force as an equivalent variance in the $c$ parameter introduced in the earlier, single-parameter sections.  We rename $c$ as $c_0$, which corresponds to the equilibrium displacement of the pendulum mass in the absence of the external force, and we add the parameter $c_1$ to account for a linear drift in the equilibrium displacement, which commonly occurs as the torsion fibre relaxes.

\subsection{Modulated Signal and Thermal Noise}

\begin{figure}
\includegraphics[width=0.45\textwidth]{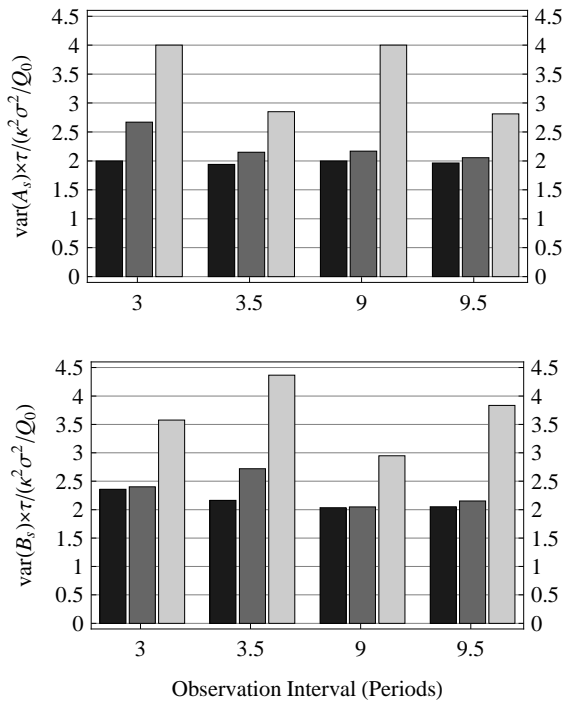}
\caption{Comparison of the estimator variances for cosine (top) and sine (bottom) signal amplitudes for the multiparameter fit (see (\ref{parameterizationeq})) using the optimal thermal-noise estimator, the \ew\ estimator, and the trigonometric estimator in the presence of thermal noise.  The grayscale convention is the optimal estimator (black), the \ew\ estimator (dark gray), and the trigonometric estimator (light gray).}
\label{multifigure1}
\end{figure}

With the mathematical tools presented in the previous section we calculate the variance of the optimal, multi-parameter estimators for the signal parameters $a_s$ and $b_s$ when accounting for thermal noise exhibited by the oscillator.  We compare those variances with the variances of the estimators from the trigonometric and \ew\ approaches.  The results of these calculations for 3, 3.5, 9 and 9.5 periods are given in Figure \ref{multifigure1}.  

We see in Figure \ref{multifigure1} that modulating the signal causes the optimal thermal noise variance to be roughly a factor of two larger than the variance for the constant force estimator.  This factor of two could be recovered if the force were modulated as a square wave instead of a sinusoid---a technique  occasionally implemented \citep{hoskins85,cowsik90}.  Regardless, sinusoidal modulation offers the advantage of simultaneously measuring both quadrature components of the signal.  Thus, if one requires knowledge of both of these parameters, then no measurement time is lost.  Moreover, sinusoidal modulation of the signal can help reject gravitational systematics with higher azimuthal symmetries (such as $\cos(3 \omega_s t)$ or $\sin(3 \omega_s t)$).

When fitting these parameters optimally for thermal noise, the variances of the several sine amplitude estimators are larger than those of the cosine amplitudes.  This effect is due to the greater functional overlap between the $b_s$ and $c_1$ parameters  compared to that between $a_s$ and $c_0$.  The same effects arise for the case of the optimal white noise fit to these same four parameters.  Indeed, the variances of the optimal thermal-noise estimators are identical to those of the optimal white noise estimators without the factor of 9/5 in the expressions for $a_s$ and $b_s$, and with the ``equivalent'' white noise power defined as in (\ref{whitenormalize}) of Section \ref{whitenoise}.

To see why this is so, let us first denote the case of an optimal white estimator in the presence of white noise with the subscript ``$I$'' and the case of an optimal thermal-noise estimator in the presence of thermal noise with the subscript ``$II$''.  The component functions of the design vector for the optimal white-noise fit are then, $\B{q}_{I}(t) = \{\cos (\omega_s t)$, $\sin(\omega_s t)$, $1$, $t$\}.  For the optimal thermal-noise fit (factors of 9/5 included) they are $\B{q}_{II}(t) = \{9/5 \cos (\omega_s t)$, $9/5 \sin(\omega_s t)$, $1$, $t$\}.  The stated equivalence arises because the four functions of $\B{q}_{II}(t)$ in the driving-force basis are directly proportional to the four functions of $\B{q}_{I}(t)$ in the displacement basis.  That is, to zero-order in 1/Q,
\begin{equation}
\Omega\left[\B{q}_{II}(t)\right] = \kappa \B{q}_{I}(t).
\end{equation}
Since all four components of $\B{q}_{II}(t)$ are non-zero, it can be shown that the optimal estimators for thermal noise must be force-only, and that they may be calculated in the driving-force basis.  Specifically there would be an analog to (\ref{whiteeop}) in Section \ref{multiparams} with $\matx{e}_{\hat{p}}^{\op}$ replaced with $\matx{y}_{\hat{p}}^{\op}$ and with $\B{q}(t)$ replaced with $\Omega\left[\B{q}(t)\right]$.  From this, one then gets
\begin{equation}
\matx{y}_{\hat{p}II}^{\op} = \matx{e}_{\hat{p}I}^{\op}/ \kappa.
\end{equation}
From (\ref{whitenormalize}), one notes a distinguishing factor of $\kappa^2$ between the driving-force PSD of case $II$, $4k_{\bsub}T\xi$, and the displacement PSD of case $I$.  When calculating the variances in the parameter estimates, all of these factors of $\kappa$ cancel, yielding identical values for both the optimal white estimator with white noise and the optimal thermal estimator with thermal noise.

For the \ew\ estimators, pre-filtering the data leads one to ask whether the signal amplitude is thereby reduced by a factor of two.  In fact, it is; however, the noise is also filtered.  For a long data duration the FED is narrowbanded around the signal frequency and the ratio of the attenuation in the signal and the attenuation in the noise is very nearly unity.  In this manner, the \ew\ estimator approaches the optimal just as it did for the static case presented in Section \ref{secthreea}.  In essence, this argument is the Fourier representation of that discussion.

Unlike the \ew\ estimators, the trigonometric estimators do not converge to the optimal for a modulated signal.  The narrowbanding argument from the previous paragraph does not apply for the trigonometric estimators because a significant contribution from the resonance frequency is required in order to match the necessary boundary conditions of a force-only estimator---that it have smooth transitions to zero.  Consequently, the effects on the variance of matching these boundary conditions are not localized to the beginning and end of the estimating function, and thus their contributions do not diminish to zero with longer duration measurements.

The DC-signal trigonometric estimators discussed in Section \ref{comparisonsec} may provide a good qualitative understanding of the modulated-signal trigonometric estimators in this section.  However, a comparison of Figures \ref{variancecomp1} and \ref{multifigure1} show important quantitative differences in the estimator variances between the modulated and static cases.  For example, a common assumption that the DC signal is a special case of the cosine component of the modulated signal proves incorrect as the variance of the cosine amplitude is larger than optimal by a factor of two instead the factor of $3/2$ for the static case of Section \ref{comparisonsec}.  Thus, a specific calculation of the estimator variances is required to achieve reliable numerical results for a modulated signal.  

\subsection{Modulated Signal and White Noise}

\begin{figure}
\includegraphics[width=0.45\textwidth]{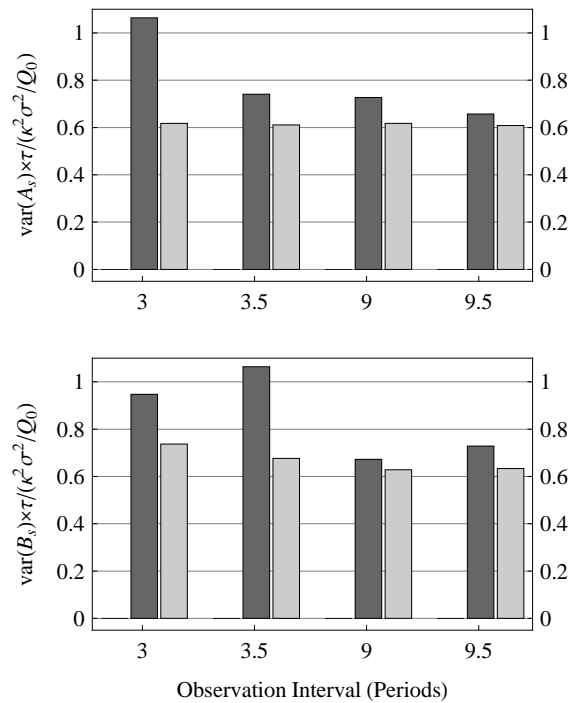}
\caption{Comparison of the estimator variances for cosine (top) and sine (bottom) signal amplitudes for the multiparameter fit (see Eqn. \ref{parameterizationeq}) using the \ew\ estimator and the trigonometric estimator in the presence of white noise.  The grayscale convention is the same as in Figure \ref{variancecomp1}.}
\label{multifigure2}
\end{figure}

The white-noise counterparts to the thermal-noise results in Figure \ref{multifigure1} are shown in Figure \ref{multifigure2}.  The estimator variances in the presence of white noise are generally a factor of $(5/9)^2$ smaller than the corresponding estimators in the presence of thermal noise.  This follows because the displacement basis signal becomes 9/5 larger while the additive noise remains the same.  For thermal noise, the response of the oscillator causes both signal and noise to increase by the same factor, leaving the signal-to-noise ratio unchanged.

The trigonometric estimators are nearly optimal in the presence of white noise because they are, in essence, optimal white-noise fits but with two additional terms, $a_0 \cos(\omega_0 t)$ and $b_0 \sin(\omega_0 t)$.  The differences from the optimal case are due to the functional overlap of $a_0$ and $b_0$ with the other four parameters.  Since these overlaps tend to decrease as the data duration increases, the trigonometric estimators converge to the optimal.

\begin{figure}
\includegraphics[width=0.45\textwidth]{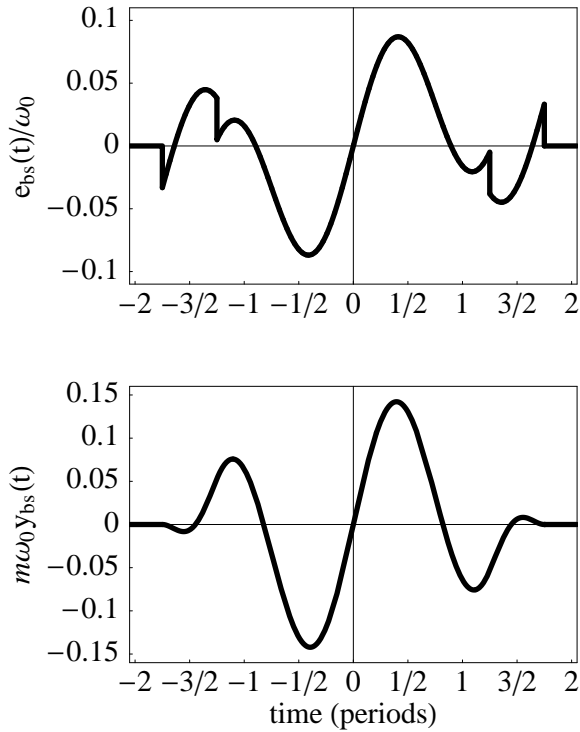}
\caption{Comparison of the $e_p$ (top) and $y_p$ (bottom) for the \ew\ estimator for the 3.5-period sine amplitude.  The jump discontinuities that are present in the $e_p$ give the \ew\ estimators larger variance than the trigonometric estimators in the presence of white displacement noise.  For thermal noise the same does not occur because, in the basis where the thermal driving force is white and the $y_p$ is used, there are no such discontinuities.}
\label{compareeandy}
\end{figure}

When compared with the optimal estimators for either noise process, the \ew\ estimators are much closer to optimal for thermal noise than they are for white noise.  The reason for this behaviour is subtle, as the arguments for high performance of \ew\ estimators in the presence of thermal noise---that the central portion of the $y_p$'s approximate matched filters well and that only the first and last quarter periods deviate significantly from optimal---would seem to apply just as well to the estimating functions in the displacement basis that are used to calculate the white noise variances.  The reason that the \ew\ estimators perform less well in the presence of white noise becomes clear in Figure \ref{compareeandy}.  By comparing the $e_p$ and $y_p$ for the sine amplitude in the 3.5-period trigonometric estimator we see that, by construction, the $e_p$ has jump discontinuities.  The $y_p$, on the other hand, is obtained through a convolution of the $e_p$ (see (\ref{yfrome}) in Section \ref{oneperiod}) and therefore must be a continuous function.  The $e_p$, with its jump discontinuities, is thus more free to deviate from an optimal matched filter.

\subsection{Mixed Noise}

What about mixed noise and multiparameter fits?  For a modulated signal and with the addition of the polynomial parameters $c_0$ and $c_1$, we can identify the ratios of thermal to white noise where the \ew\ estimation technique gives equal variances to the trigonometric approach as done in Section \ref{whitenoise}.  These results are given in table \ref{noisemixmulti}.

\begin{table}
\caption{\label{noisemixmulti}Noise mixing ratios for various sample durations where the trigonometric estimators become superior to the \ew\ estimators for the four-parameter model discussed.  The \ew\ estimators are superior when the white noise contribution is less than the stated amount.}
\begin{center}
\item[]
\begin{tabular}{lcccc}
\hline
Periods              & 3    & 3.5  & 9    & 9.5  \\
\hline \hline
White Noise Fraction & 79\% & 82\% & 95\% & 94\% \\
\hline
\end{tabular}
\end{center}
\end{table}

For a measurement scheme in which white noise dominates, the experimenter should likely consider increasing the modulation frequency closer to resonance in an effort to increase the signal-to-noise ratio (the signal-to-noise ratio for thermal noise is independent of the modulation frequency).  In such a situation a resonant method would be suitable.  With some modification to accommodate non-linear fits, the techniques developed in this article can be applied to resonant and other large-oscillation-amplitude detection schemes \citep{cowsik90,boynton07} though a detailed exploration of that topic lies beyond the scope of this work.

\subsection{1/f Thermal Noise}

For modern torsion balance experiments that are conducted at high vacuum, the predominant damping/noise mechanism is the fibre dissipating energy rather than residual gas in the vacuum chamber.  \citet{saulson90} states that for this scenario a $1/f$ profile is a better approximation to the driving-force PSD than the white driving force that is characteristic of a classical dashpot (although $1/f$ is the common term for this noise process, we use $1/\nu$ in our formulas in order to avoid confusion with filter functions).  Mathematically, such a spectrum is less tractable, so there is no simple method for finding the optimal estimating functions as for the oscillator with dashpot damping.

Nevertheless, we can still calculate and compare the variances of our estimating functions.  We compute these quantities in the driving-force basis since it is still easier than in the displacement basis, but we must now numerically integrate the product of the PSD and the FED (see (\ref{varequation})).  We first note that this revised damping of the oscillator is frequency dependent with a functional form similar to that of the driving force:
\begin{equation}
\xi(\nu) = \frac{\xi_0 \nu_0}{\nu}.
\end{equation}
where $\nu_0 = \omega_0/(2\pi)$ is the resonance frequency (in Hz) and $\xi_0$ is the velocity damping coefficient that corresponds to the resonance frequency.  This form constrains the observed damping at the resonance frequency to be the same for both $1/f$ thermal noise and for dashpot thermal noise.  The variance in the parameter estimators for $1/f$ noise is then
\begin{equation}
\begin{split}
\text{var}(\hat{P}_{1/f}) &= \frac{1}{2}\halfint F^2[y_{\hat{p}}] S[\Omega[\delta X_{1/f}]] d\nu \\
&= \frac{1}{2}\halfint F^2[y_{\hat{p}}] \left[4k_{\bsub} T \xi (\nu) \right] d\nu \\
&= 2k_{\bsub}T\xi_0\halfint F^2[y_{\hat{p}}]\frac{\nu_0}{\nu}d\nu.
\end{split}
\end{equation}
Here the $1/f$ subscript signifies that the driving-force PSD has a $1/f$ profile.

\begin{figure}
\includegraphics[width=0.45\textwidth]{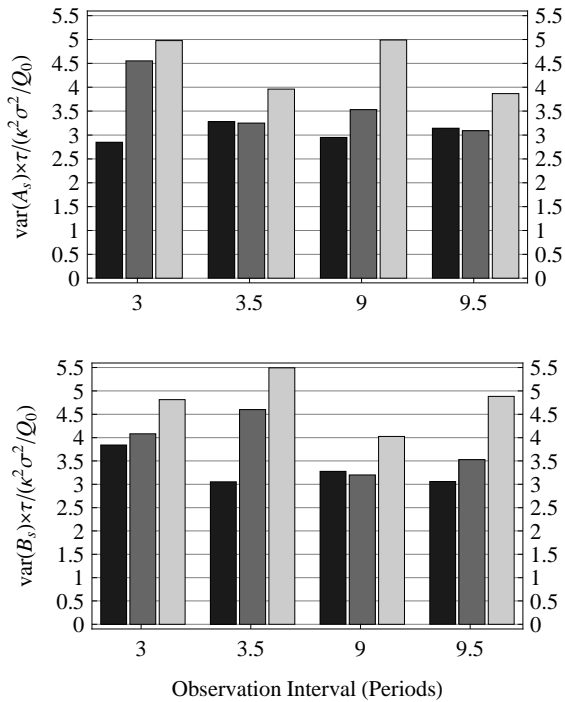}
\caption{Comparison of the estimator variances for cosine (top) and sine (bottom) signal amplitudes for the multiparameter fit (see (\ref{parameterizationeq})) using the optimal thermal-noise estimator, \ew\ estimator and the trigonometric estimator in the presence of $1/f$ noise.  The grayscale convention is the same as in Figure \ref{variancecomp1}.}
\label{oneoverffigure}
\end{figure}

The zero-frequency PSD singularity for $1/f$ noise requires the FED of the parameter estimating function $F^2[y_p]$ to have a corresponding zero in order for the variance to be well defined.  That is, $y_p$ must be orthogonal to a constant force signal.  The estimating functions for $a_s$ and $b_s$ are orthogonal to the constant signal parameterized by $c_0$, and since they are force-only estimators, it can be shown that their corresponding $y_p$'s are also orthogonal to a constant.  Thus, the variance of the estimators for $a_s$ and $b_s$ remain finite.  The variances for all three methods---optimal thermal, \ew, and trigonometric---are shown in Figure \ref{oneoverffigure} for different observation intervals.  Once again, the signal frequency $\omega_s = 2 \omega_0 / 3$.

Comparing variance in the optimal thermal-noise estimator for the case of $1/f$ thermal noise (Figure \ref{oneoverffigure}) with dashpot thermal noise (Figure \ref{multifigure1}) we see that the variance with $1/f$ noise is larger by about 50\%.  This is largely because that noise is greater by 50\% at the chosen signal frequency.  We also see from Figure \ref{oneoverffigure} that the optimal thermal-noise estimator is slightly inferior to the \ew\ estimator for some values of the sample duration---showing that the optimal thermal-noise estimator is not the true optimal estimator for $1/f$ thermal noise, as expected.  However, a comparison of the results for $1/f$ thermal noise in Figure \ref{oneoverffigure} with the results of both dashpot thermal noise in Figure \ref{multifigure1} and white displacment noise in Figure \ref{multifigure2}, indicates that dashpot thermal noise is the better model to guide the development of an estimator for $1/f$ thermal noise.

\section{Discussion}

The topics discussed in this paper illustrate the differences between two selected parameter estimation techniques relative to the optimal technique for thermal noise dominated experiments.  A major simplification in the study and application of parameter estimation for this case results from working in the basis of the thermal driving force where the PSD of the noise is white.  Equations (\ref{eqofmotiontrans}) and (\ref{yfrome}) show how to transform between the displacement basis and the driving-force basis.

In the driving-force basis, analysis practices that are commonly applied to data exhibiting white displacement noise can be utilized for estimation in the presence of thermal noise.  Working in that basis allows better insight to and understanding of the results from various approximate methods that experimentalists generally employ.  We also showed that in the driving-force basis, one can more readily construct estimators, such as the \mix\ estimator, that perform well for the case where the noise is a mixture of white and thermal.

Additionally, working in the driving-force basis ensures that a parameter estimator is insensitive to the effects of the resonance peak.  This feature also provides immunity to transient oscillations induced by disturbances that occur prior to the data sample, as discussed in Paper~I.  A further advantage, for modulated signals, is that the signal frequency does not have to be commensurate with the natural oscillation frequency.  In fact, there can be two or more signal frequencies that need not be commensurate with each other.

While the approach we outline may be applied generally, the interesting calculations can be truncated to leading order in $1/Q$ without losing value to an experimentalist.  Specifically, meaningful results are obtained---and with greater facility---using the undamped approximation for the signal and estimating functions.  Besides making analytic calculations simpler, working in the driving-force basis makes leading-order numerical solutions more robust against artificial influence of power from the resonance peak.

For a real torsion pendulum experiment, the superposition of several noise processes is a challenge that requires one to adopt a data analysis scheme that performs well under a variety of circumstances, while necessarily being optimal in none.  We have shown that the \ew\ estimation technique, formerly used by the E\"ot-Wash group, is not only well behaved in the presence of white displacement noise, thermal noise, and $1/f$ noise but approaches the performance of the optimal for both white and thermal noise as the duration of the observations increase.  For $1/f$ noise, the \ew\ estimator also performs well relative to the other techniques studied in this paper, though the singular nature of the $1/f$ noise spectrum renders identifying its corresponding optimal estimators problematic.  By using the dashpot thermal noise results as a guide, one can correctly infer that the \ew\ estimator is superior to the trigonometric estimator for $1/f$ thermal noise; whereas if one had used the white noise results as a guide, one might erroneously assume the trigonometric estimator to be superior.

Even with the mathematical tools developed here and in Paper~I, there remain several issues that are important to a modern experimental research program that lie beyond the scope of this paper.  One obvious example is our \textit{a priori} assumption regarding knowledge of the oscillation frequency and quality factor of the pendulum.  One would prefer to identify these quantities from the same data used to estimate the other parameters of the system---especially if the experiment uses a pendulum that undergoes large amplitude oscillations.  Moreover, for extension to the case of large amplitude oscillations, the nonlinear nature of both the oscillation frequency and the decay constant require an additional generalization of the techniques discussed here.  This is even true of the linear signal parameters of non-resonant, higher harmonic methods, such as the treatment of the second harmonic we employ in our own experiments because signal-to-noise ratio considerations require operation with large-amplitude oscillations \citep{boynton00}.  We leave for further investigation the extension of techniques introduced here to the optimal estimation of multiple, nonlinear parameters and its implications for torsion pendulum experiments.

\acknowledgements
We gratefully acknowledge NSF Grants PHY- 0244762, -060692,  and -071923 for
support of this work.  JHS recognizes the generous support of the Brinson Foundation and Fermilab under DOE contract No. DE-AC02-07CH11359.

% Create the reference section using BibTeX:

\end{document}